%% file: main.tex
\documentclass{article}
\usepackage{graphicx} 
\usepackage[a4paper, left=3cm, right=3cm, top=3cm, bottom=3cm]{geometry}
\usepackage{textgreek} 
\usepackage[utf8]{inputenc}
\usepackage[backend=biber, style=apa, natbib=true]{biblatex}
\usepackage{float}
\usepackage{tabularray}
\usepackage{codehigh}
\usepackage[normalem]{ulem}
\usepackage[hidelinks]{hyperref}
\usepackage{setspace}
\usepackage[T1]{fontenc}
\usepackage[utf8]{inputenc}
\usepackage{booktabs}     
\usepackage{longtable}   
\usepackage{siunitx}     
\newcommand{\sym}[1]{\textsuperscript{#1}} 
\begin{document}

\begin{titlepage}
    \thispagestyle{empty}
    \centering

    {\scshape\LARGE Potsdam University \par}
    \vspace{1cm}

    {\Huge\bfseries
    Gendered Responses to Subtle Social Pressure: Experimental Evidence from Survey Results\footnotemark\par}
    \vspace{1cm}

    {\Large\itshape Sevgi Çolak\par}
    \vspace{1.5cm}

    \begin{abstract}
        \onehalfspacing
       
        This study analyzes whether subtle variations in the survey questionnaire phrasing influence participant engagement and whether these effects differ by gender. Building on theories of social pressure and politeness norms, it is hypothesized that presumptive phrasing would reduce engagement compared to appreciative phrasing and baseline phrasing (H1), and this effect would be more pronounced among women (H2). Mixed-effects regression models showed no significant treatment effects on any outcome and no evidence of gender moderation for 164 participants and 492 observations. The only robust finding was a small negative baseline sentiment across all participants, independent of any treatment or gender. The findings contribute to refining theoretical expectations about the conditions in which linguistic framing and gender norms shape behaviour.

        \medskip
        \textbf{Keywords:} social pressure, survey design, politeness framing, gratitude language, gender differences, response engagement, response effort, behavioral economics.
    \end{abstract}

    \vspace{2cm}

    Supervisor: Prof. Dr. Lisa V. Bruttel \par
    \vspace{1cm}

    {\large Berlin, August 2025 \par}

    \vfill

    \footnotetext{\href{https://osf.io/xsq29}{Preregistration available here, "https://osf.io/xsq29"}.}
    
\end{titlepage}

\section{Introduction}

Social pressure has a profound impact on human communication, influencing how people interact in various contexts. In survey settings, small social cues could influence participants' willingness to engage and the effort they put into their responses. Understanding how such cues affect individuals' interaction with a survey is crucial for improving the quality and accuracy of survey-based research.

This paper will focus on two main questions, which are;

1) Does presumptive phrasing reduce response engagement on survey questions, as measured by criteria such as response length, sentiment, and time spent, when compared to appreciative language? 

2) Does gender influence the effect of social pressure on survey responses, with female participants showing a greater sensitivity to presumptive language than males?

To expand further, this study examines how subtle social pressure cues influence responses to survey questions. In addition to the presumptive and appreciative experimental groups, a control group is also included to compare the effects of social pressure cues to the absence of any cue. This will help isolate the effects of the phrasing itself and aims to remove any external factors that may influence responses.

Gender is considered at all stages of the analysis, and the evaluation of gender as an independent variable is performed using multivariate regression analysis. This ensures a thorough analysis of possible binary gender differences in survey engagement. 

The results show that neither presumptive nor appreciative phrasing yielded measurable differences in engagement with the survey. Similarly, gender did not moderate the effect of phrasing, as men and women responded comparably across all conditions. The only significant finding was a slightly negative baseline sentiment, independent of treatment or gender. Subtle differences in closing phrases might not meaningfully influence participant behaviour in short survey tasks, as suggested by the null results.

\section{Literature Review}

The impact of presumptive versus appreciative phrasing on compliance has been directly tested in recent experimental research. \cite{bruttel2022thanks} showed that the phrase ``thanks in advance'' significantly reduced compliance effort in survey-like tasks, as participants provided shorter and less elaborate responses. This finding directly motivates H1, which predicts that presumptive phrasing diminishes engagement compared to appreciative alternatives. In contrast, \cite{andreoni2011asking} demonstrated that expressions of gratitude such as ``thank you'' increased generosity in a dictator game, suggesting that appreciation can foster prosocial responses. Taken together, these studies establish the theoretical basis for distinguishing between presumptive and appreciative language in surveys.  

Gendered responses to subtle cues further highlight the importance of examining moderation effects. \cite{bruttel2018gender} found that women were more responsive to freedom-of-choice framing, whereas men reacted more strongly to responsibility framing, indicating that linguistic nuances interact with gender-specific expectations. Similarly, \cite{dolinska2006command} observed that the effectiveness of polite versus direct request strategies varied with the gender of the target. Extending these findings, \cite{becker2022gender} documented that women display consistently higher survey participation rates than men, partly due to altruistic and socialized helping tendencies. These studies jointly support H2, which declares that phrasing effects will be more pronounced among female respondents.  

Broader evidence reinforces the role of linguistic framing in compliance. Large-scale text analyses show that gratitude and reciprocity cues increase cooperation. \cite{althoff2014favor} reported that online favor requests containing gratitude and pay-it-forward signals were more successful, while \cite{mitra2014language} found that crowdfunding campaigns that used appreciative and inclusive phrasing attracted more support. These findings suggest that appreciation is a robust predictor of compliance across contexts. Finally, survey methodology research has established that structural factors also affect engagement. For instance, \cite{galesic2009lenght} showed that longer surveys reduce participation and degrade response quality, underscoring the challenge of maintaining engagement and the potential value of linguistic strategies to offset attrition.

\section{Hypotheses}

This study is guided by two primary hypotheses, building on research on social pressure, compliance, and gendered communication. The first hypothesis examines the overall effect of phrasing, while the second tests gender as a moderator. 

\textbf{H1}: Presumptive phrasing will reduce response engagement compared to appreciative phrasing and the control condition.  

\textbf{H2}: The difference in response engagement across phrasing conditions will be more pronounced among female participants than male participants.  

These hypotheses draw on prior findings that women are generally more attentive to social cues and more reactive to violations of autonomy or politeness norms. In particular, \cite{bruttel2022thanks} shows that presumptive language can provoke negative reactions, especially among women in compliance settings. Accordingly, female participants are expected to respond more negatively to presumptive phrasing than men. 

While H1 considers the general effect of phrasing, H2 represents the central focus of this study, as it addresses potential gender differences in sensitivity to subtle social pressure with important implications for survey design and behavioral research. Existing evidence underscores this possibility: women often show higher participation and willingness to respond when incentives are present \citep{singer2008incentives}, yet they may also react more negatively to pressure cues \citep{bruttel2022thanks}. Relatedly, women tend to maintain slightly higher response rates in both telephone and online surveys over time, even as overall participation has declined \citep{curtin2000response}. Together, these findings motivate testing whether gender moderates the effects of subtle phrasing manipulations.

\section{Methodology and Experimental Design}

The experimental manipulation concerned the closing wording of survey instructions in a within-subjects design with three conditions. The Presumptive language phrasing read “Thank you in advance for sharing your thoughts!”\footnote{German: ``Vielen Dank im Voraus für das Teilen Ihrer Überlegungen!''}, the Appreciative language phrasing saw “We would be grateful if you could share your thoughts with us.”\footnote{German: ``Wir wären Ihnen dankbar, wenn Sie Ihre Überlegungen mit uns teilen.''}, and for the baseline statement, participants received no additional phrasing. This design enables direct comparison of how subtle linguistic differences shape response behavior. Gratitude is most effective when perceived as sincere appreciation, but thanking in advance may be interpreted as insincere or manipulative (\cite{dwyer2015gratitude, carey1976reinforcement}).

Presumptive phrasing assumes engagement and may create obligation, potentially provoking resistance, whereas appreciative phrasing conveys hope without presumption, acknowledging the respondent’s agency. The control group provides a neutral baseline. The primary contrast of interest is between presumptive phrasing and control (H1), with presumptive versus appreciative phrasing examined secondarily. Gender is central as a moderator (H2), testing whether women respond more strongly to presumptive cues.

Engagement was assessed through three indicators: response length (word count), response time (seconds), and sentiment of responses\footnote{Sentiment was measured using the SentiWS (SentimentWortschatz) lexicon \citep{remus2010sentiws}, which assigns polarity scores to German words based on their emotional tone.}. Longer answers and greater time investment were taken as proxies for deeper engagement, while sentiment captured emotional tone. Although sentiment can reflect both thoughtful engagement and social desirability, it remains a meaningful indicator of participant involvement. Gender, self-reported at the end of the survey, has been considered binary, meaning male or female, for analysis. While participants could also select “other,” this category was excluded due to small numbers. 

An initial composite index combining z-scores for length, time, and sentiment failed reliability checks (Cronbach’s $\alpha < 0.70$). Instead, a sub-index combining length and time was used, meeting the $\alpha > 0.70$ threshold. This provided a more stable measure of engagement while reducing noise.

\subsection{Study Plan and Design}

The study has taken place in a laboratory experiment at the Potsdam Laboratory for Economic Experiments (PLEx). Between behavioral economics games, participants answered four open-ended evaluation questions, each randomly assigned a different phrasing condition: one presumptive, one appreciative, and two control. Randomization ensured order balance and comparability across conditions. Because the manipulation was not disclosed, participants remained blind to the treatment.

A priori power analysis (G*Power 3.1.9.7) indicated a minimum of 164 participants for detecting small effects ($dz = 0.2$, $\alpha = 0.05$, power $= 0.80$) using Wilcoxon signed-rank tests. This sample was sufficient to test both H1 and H2 while allowing within-subject comparisons and subgroup analyses by gender. The matched-pairs structure increased efficiency by controlling for individual heterogeneity.

The data used in this study were originally collected for a laboratory experiment that also forms the basis of another, as yet unpublished, study entitled "A Simple Measure of Strategic Uncertainty Aversion". While the two projects draw on the same dataset, this paper focuses exclusively on survey responses and engagement outcomes. Participants who failed to complete the survey or gave nonsensical responses were excluded. Outliers in response length and time (values beyond three standard deviations) were assessed in sensitivity checks but retained in the main analysis unless clearly distorting results. All exclusions and their impact on sample size were transparently reported.

\section{Results}

The analysis consists of two phases. Initially, descriptive plots are used to show how sentiment and engagement vary among binary gender groups and treatment conditions. This allows for a visual examination of whether any noticeable trends between the male and female participants emerge. Second, the effects of gender and treatment phrasing on various outcome measures, including engagement index, duration, response length, and sentiment, are statistically assessed using mixed-effects regressions.

\subsection{Binary Gender Analysis}

\begin{figure}[h]
  \centering
  \tiny
  \includegraphics[width=\linewidth]{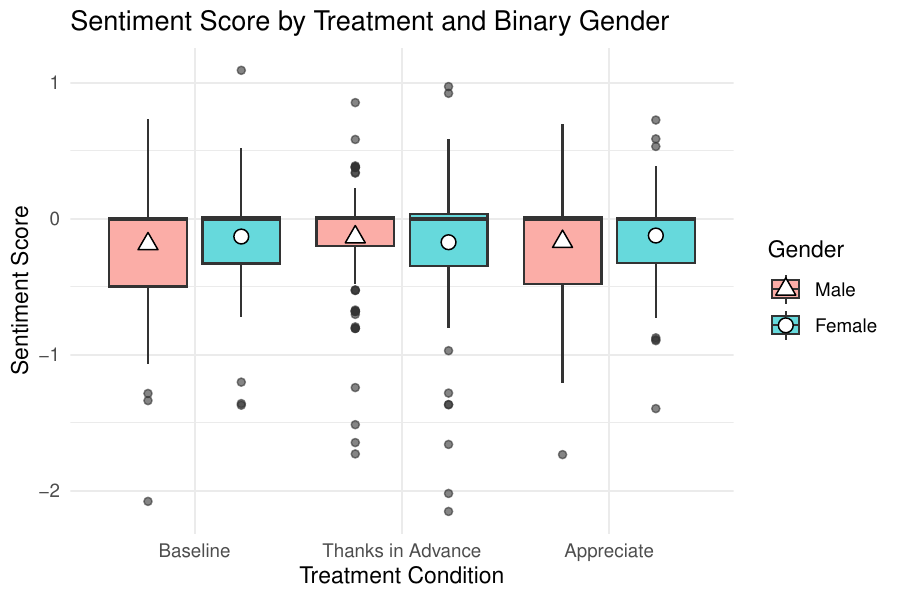}
  \caption{Sentiment score by treatment and binary gender.}
  \label{fig:sentiment-binary}
\end{figure}

\begin{figure}[H]
  \centering
  \tiny
  \includegraphics[width=\linewidth]{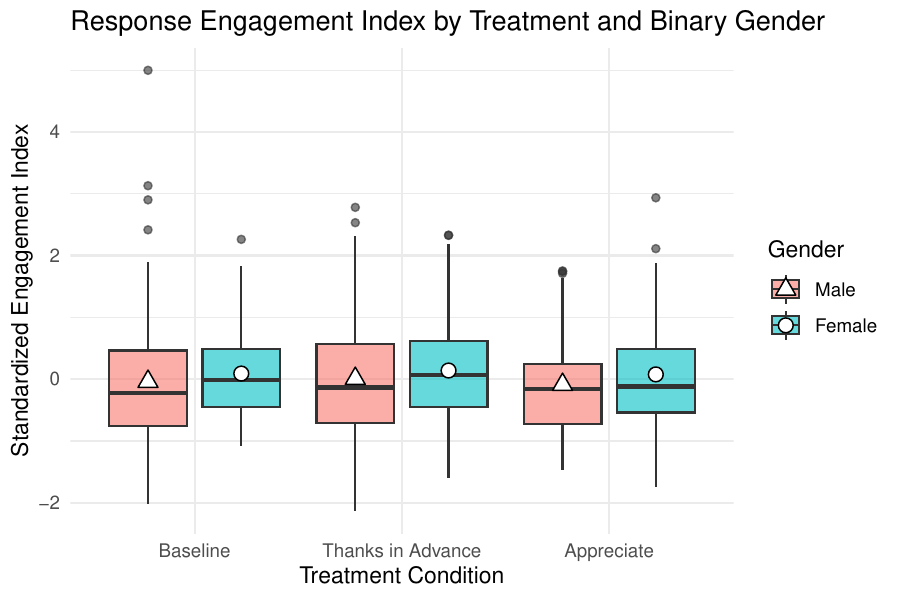}
  \caption{Response engagement index by treatment and binary gender.}
  \label{fig:engagement-binary}
\end{figure}

Figures [\ref{fig:sentiment-binary}] and [\ref{fig:engagement-binary}] present the distribution of sentiment scores and engagement indices across treatment conditions, disaggregated by gender.

In the case of sentiment (Figure [\ref{fig:sentiment-binary}]), responses from both female and male participants are concentrated around zero across all three phrasing conditions. Median values are nearly identical, and the interquartile ranges overlap almost completely, with only minor variation in the tails. This pattern indicates that participants generally expressed neutral to slightly negative sentiment regardless of whether the request closed with a presumptive phrase (``Thanks in Advance''), an appreciative phrase (``Appreciate''), or the baseline control. Importantly, no visible divergence appears between genders, suggesting that the linguistic manipulations did not alter the affective tone of responses. The overall tendency toward mild negativity may reflect participants’ attitudes toward the task itself (e.g., effort required, lack of intrinsic motivation) rather than treatment-induced differences.

For engagement (Figure [\ref{fig:engagement-binary}]), median scores again cluster near zero, and the spread of the distributions is similar across all groups. Both female and male participants display comparable variability, and no systematic shift is evident across treatment conditions. The absence of visible separation between distributions implies that neither presumptive nor appreciative phrasing altered how much engagement participants invested, as measured by length and time. Moreover, the parallel distributions across genders indicate that women did not respond more sensitively to phrasing than men, contrary to expectations drawn from prior literature on gendered communication.

Taken together, the descriptive evidence reinforces the main statistical results: gender does not moderate the relationship between phrasing manipulations and participant behavior. Instead, both sentiment and engagement appear to be driven by broader situational or dispositional factors, with little indication that subtle wording choices shifted outcomes in this context.

\subsection{Regression Results}

The effects of treatment phrasing and gender were examined on four outcomes: engagement index, duration, response length, and sentiment score. Table [1] reports the mixed-effects regression results. In each model, the intercept represents the baseline condition for male participants. 

The Engagement Index ("Engagement" on the table) captures the overall quality of engagement with the reflection task, the Duration variable measures the time spent on the task in seconds, Response Length ("Length") reflects the number of words written, and the Sentiment Score ("Sentiment") indicates the average emotional tone of the response. The results show that neither presumptive phrasing ("TIA") nor appreciative phrasing ("Apprec.") significantly alters engagement, time spent, length, or sentiment compared to the baseline phrasing. Female participants, on average, tend to write somewhat longer responses and show slightly higher engagement, but these differences are not statistically significant. No meaningful interaction effects between treatment and gender emerge, indicating that subtle wording variations do not differentially influence male versus female participants. Overall, the findings suggest that response engagement and sentiment stay stable across treatments and genders.

\input{regression}

Treatment effects are consistently small and non-significant. When compared to the baseline phrasing, neither presumptive phrasing ("Thanks in Advance") nor appreciative phrasing ("Appreciate") changes sentiment, time spent, engagement, or response length. For instance, there is no tangible influence because the coefficients for TIA and Apprec. across outcomes stay near zero with standard errors greater than the point estimates.

Moreover, gender effects are still minimal. On average, female respondents are slightly more engaged (+0.13 on the index) and slightly longer (+5 words), but these differences are not statistically significant. Treatment effects do not differ between men and women, as indicated by the treatment × gender interaction terms, which are insignificant and do not reach significance across all outcomes.

The participant-level intraclass correlation coefficients (ICCs) for engagement, duration, and length. Treatment effects are consistently small and non-significant. Neither presumptive phrasing (“Thanks in Advance”) nor appreciative phrasing (“Appreciate”) alters engagement, time spent, length of response, or sentiment compared to the baseline phrasing. The coefficients for TIA and Apprec. across outcomes remain close to zero with standard errors larger than the point estimates, indicating no detectable impact.

When considered collectively, the regression evidence shows a high degree of stability across conditions: neither binary gender nor closing phrase treatments significantly affect sentiment, engagement, response time, or response length.

The findings provide no support for either hypothesis. Regarding H1, compared to the baseline condition or appreciative phrasing, presumptive phrasing ("Thanks in Advance") does not lower response engagement.  Regression coefficients for both phrasing manipulations are small and statistically insignificant, and engagement scores stay close to zero for all treatments.  Regarding H2, there is no indication that treatment effects are moderated by gender.  Although the average level of engagement is slightly higher among female participants, the difference is not statistically significant, and the treatment-gender interaction terms are still insignificant.  Collectively, these findings suggest that minor phrasing changes have no effect on levels of engagement and that female participants are not more susceptible to these manipulations than male participants.

\section{Discussion and Conclusion}

\subsection{Discussion}

This study set out to examine whether subtle variations in request phrasing influence how participants respond, and whether these effects differ between men and women. Specifically, we tested whether a presumptive closing phrase (“Thanks in Advance”) would reduce response effort compared to an appreciative phrase (“Appreciate”) and a baseline phrasing (H1), and whether this effect would be more pronounced among female participants (H2).

At first glance, the null results may appear disappointing, given the expectation that small wording choices could meaningfully shape how people engage with requests. However, they also carry important implications. First, the findings highlight the limits of subtle linguistic manipulations. While prior research has shown that wording can matter in contexts such as compliance, persuasion, or survey participation, our study suggests that not all forms of polite phrasing reliably influence behavior. The distinction between presumptive and appreciative closings may be too minor to alter how participants actually respond, especially in short written tasks where cognitive and social stakes are relatively low.

These findings can be situated within the broader literature on politeness, compliance, and gendered communication. Prior work has shown mixed results: while \cite{andreoni2011asking} and \cite{althoff2014favor} demonstrated that gratitude and appreciation can increase generosity and compliance, \cite{bruttel2022thanks} reported that the presumptive phrase ``thanks in advance'' reduced compliance effort, and \cite{gasper2022politeness} found that politeness mattered more for enabling refusals than for increasing compliance. Our results are consistent with this more cautious line of evidence, suggesting that subtle linguistic changes, particularly between two polite variants, may not reliably shift behavior in low-stakes survey contexts. By documenting null effects, our study helps refine the boundary conditions under which wording differences influence compliance and when gender differences in communication are likely to emerge.

An important limitation to consider is whether participants actually paid sufficient attention to the closing phrases. Since the manipulations were subtle and embedded at the end of otherwise routine survey instructions, it is possible that many respondents overlooked them or did not process them deeply enough to influence their behavior. In this case, our design may have reduced or even eliminated an effect that could emerge under conditions of higher visibility, such as when the phrasing is repeated or emphasized. Another possibility is that the effects reported by \cite{bruttel2022thanks} may be context-specific or less robust than initially suggested. Their findings could reflect particular features of their experimental design or sample rather than a generalizable effect of presumptive phrasing. Our null results therefore, contribute by providing an important replication check, suggesting that the negative influence of ``thanks in advance'' on compliance may not reliably emerge across all settings. An alternative possibility is that a ceiling effect was happening. Meaning, participants may already have been motivated to complete the task carefully, leaving little room for minor linguistic cues to further increase or decrease engagement. All explanations highlight that null results do not necessarily imply that phrasing has no effect, but rather that its impact may depend on attention, task clarity, and baseline motivation levels.

Second, the absence of gender differences challenges the assumption that women are uniformly more sensitive to subtle social pressures in communication. While gendered norms around politeness and relational work are well-documented in naturalistic settings, these norms did not manifest in this experimental context. This points to an important boundary condition; gender differences in communication may be more context-dependent than universal. They may emerge in higher-stakes or face-to-face interactions but remain muted in structured, low-stakes online tasks such as the one studied here.

It is almost impossible to forecast what the outcome would be if gender were treated as a continuous variable in this study. It definitely would be a more inclusive and realistic representation of people’s gender expression. Having a continuous variable would also help us to understand more nuanced outcomes, as it includes all gender expressions and does not exclude the diverse group from the analysis.

Finally, the consistently negative baseline sentiment is itself an interesting finding. It suggests that participants approached the task with a slight tendency toward negativity, which may reflect underlying attitudes about the task, the effort required, or simply a stylistic tendency in written responses. While not linked to treatment or gender, this pattern provides a useful reminder that participant responses are shaped by broader situational or dispositional factors that extend beyond experimental manipulations.

\subsection{Conclusion}

In conclusion, this study found no evidence that presumptive or appreciative phrasing affects participant engagement or sentiment, nor that women are more sensitive than men to such variations. Instead, responses remained unchanging across treatments and genders, with the only consistent pattern being a mildly negative baseline sentiment.

These findings contribute to the literature by clarifying the limits of linguistic framing effects in experimental settings. They suggest that subtle variations in closing phrases may not be powerful enough to alter behavior, especially in short survey tasks where personal involvement is limited. Moreover, the absence of gender moderation indicates that assumptions about women’s heightened responsiveness to social pressure should not be generalized across all contexts. Future work should extend to continuous measures of gender identity (see \cite{Brenoe2024}), which capture gender as a variation beyond binary categories.

Taken together, the results underscore the importance of testing when and where subtle forms of phrasing matter. Future research could explore whether stronger manipulations, higher-stakes tasks, or more socially interactive settings are needed to elicit measurable differences. By documenting these null results, our study provides valuable evidence that helps refine theoretical expectations and identify the boundary conditions under which social pressure and gendered communication norms truly shape behavior.

\appendix
\section*{Appendix}

\subsection*{Inclusive Gender Analysis}

\begin{figure}[h]
  \centering
  \tiny
  \includegraphics[width=\linewidth]{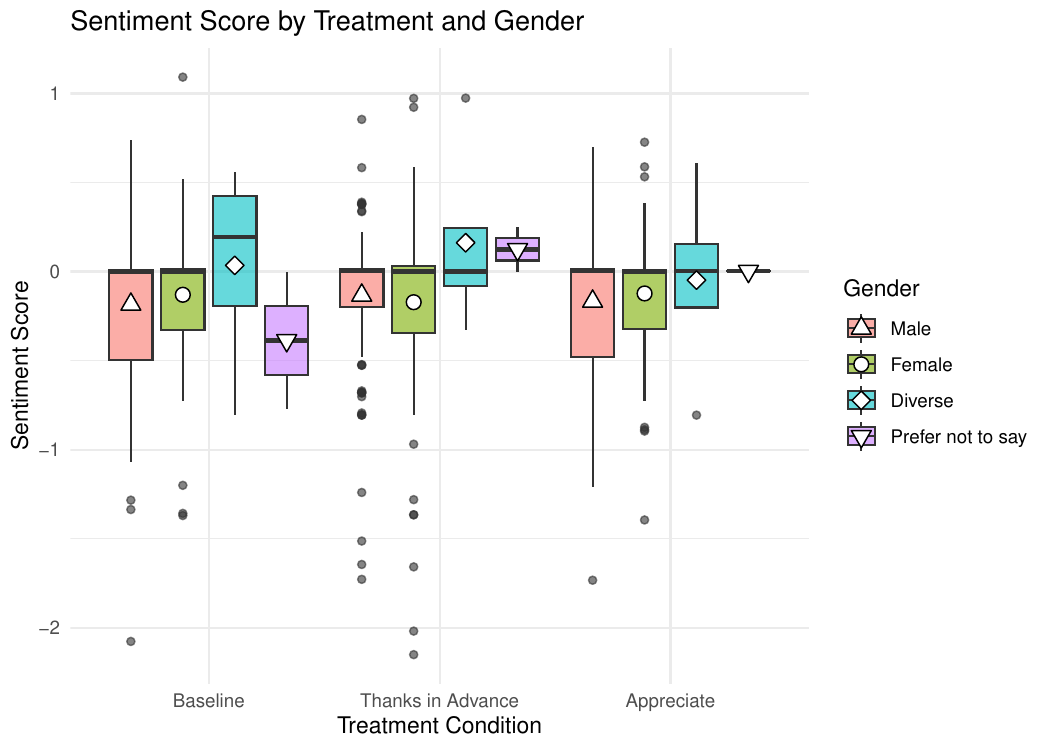}
  \caption{Sentiment score by treatment and all genders.}
  \label{fig:sentiment-all}
\end{figure}

\begin{figure}[h]
  \centering
  \tiny
  \includegraphics[width=\linewidth]{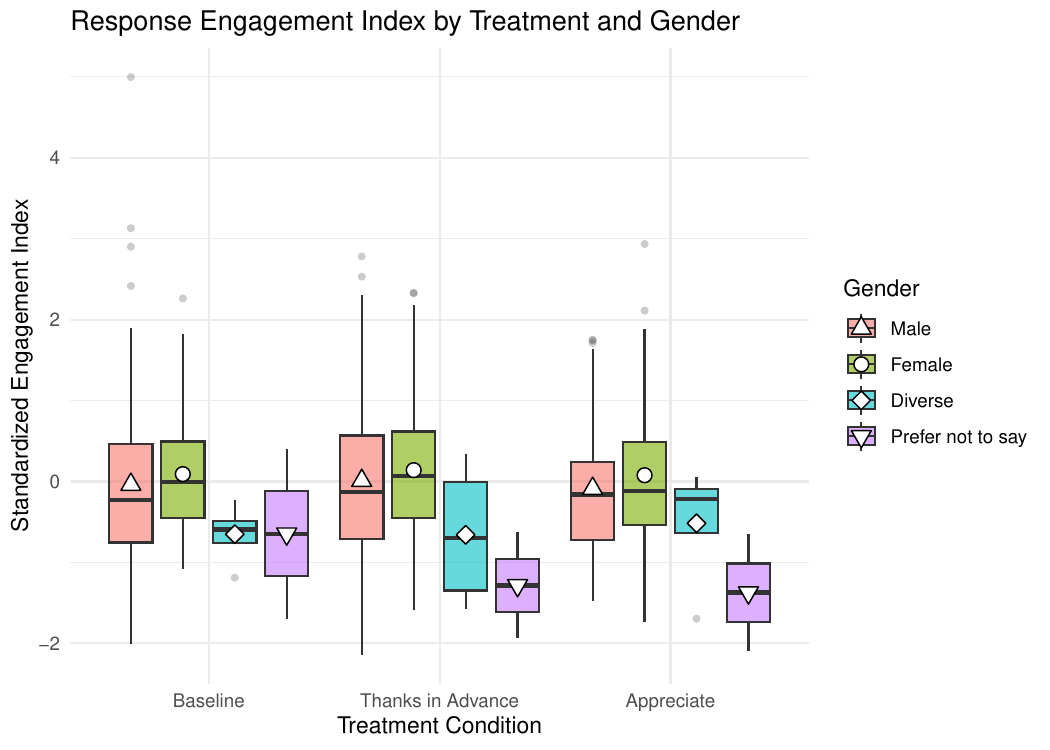}
  \caption{Response engagement index by treatment and all genders.}
  \label{fig:engagement-all}
\end{figure}

Even though the main analysis of this study considers gender as a binary variable, it is important to include a part where we look at the plots that include all four gender categories (male, female, diverse, and “prefer not to say”) (Figure [3], Figure [4]), to understand diverse gender identities better.

Figures [3] and [4] show that both sentiment and engagement scores cluster around zero across all treatments, indicating that phrasing did not produce strong emotional reactions or shifts in effort. Responses leaned slightly negative overall, suggesting limited enthusiasm regardless of condition. While the “diverse” and “prefer not to say” groups displayed greater variation, their small size makes these patterns difficult to interpret, and no treatment consistently altered engagement levels across gender groups.

In conclusion, there are no systematic differences in either the sentiment variable or the engagement index across treatment conditions or gender categories. The variation within groups is far larger than any variation between them.

\subsection*{Construction of the Engagement Index}

To capture overall response engagement, we initially constructed a composite index combining three indicators: response length (word count), response time (seconds spent on the task), and response sentiment (mean sentiment score of the text). Each measure was standardized into z-scores to ensure comparability across scales. These z-scores were then combined into a single index, with the expectation that together they would provide a holistic measure of participant engagement.

However, reliability analysis indicated that the three components did not cohere sufficiently as a single construct. Cronbach’s $\alpha$ fell below the conventional threshold of 0.70, suggesting that sentiment did not load consistently with length and time as part of a unidimensional scale. This aligns with the theoretical consideration that sentiment reflects the tone of responses rather than the level of effort or engagement per se.

In response, we tested a reduced index including only response length and response time. This two-item scale achieved satisfactory reliability (Cronbach’s $\alpha > 0.70$), indicating that these measures captured a common underlying dimension of response engagement. Using this sub-index provided a more stable and interpretable measure of engagement while minimizing noise introduced by sentiment, which was analyzed separately in the main results.

\subsection*{Acknowledgment of Tools and Software}

All analyses were conducted in \textsf{R} (version 4.5.0, “How About a Twenty-Six,” released 2025-04-11) using the following packages: 
\texttt{tidyverse}, \texttt{readxl}, \texttt{tidytext}, \texttt{tm}, \texttt{lme4}, \texttt{sjPlot}, \texttt{psych}, \texttt{dplyr}, \texttt{ggplot2}, and \texttt{modelsummary}. 
Power analyses were performed with \textsf{G*Power} (version 3.1.9.7). Sentiment analysis relied on the German lexicon \textsf{SentiWS} \cite{remus2010sentiws}. 
Generative AI assistance was provided by \textsf{ChatGPT} (OpenAI) to support language polishing and phrasing alternatives. 
Additionally, \textsf{Grammarly} was used to check grammar and style consistency. 
All interpretations, methodological decisions, and conclusions remain the sole responsibility of the author.

\printbibliography

\end{document}

%% file: regression.tex
\begin{longtable}{@{}l
  S[table-format=2.2]
  S[table-format=3.2]
  S[table-format=3.2]
  S[table-format=1.2]
@{}}
\caption{Treatment $\times$ Gender Effects Across Outcomes}
\label{tab:regression}\\

\toprule
& \multicolumn{1}{c}{Engagement}
& \multicolumn{1}{c}{Duration}
& \multicolumn{1}{c}{Length}
& \multicolumn{1}{c}{Sentiment}\\
\midrule
\endfirsthead

\toprule
& \multicolumn{1}{c}{Engagement}
& \multicolumn{1}{c}{Duration}
& \multicolumn{1}{c}{Length}
& \multicolumn{1}{c}{Sentiment}\\
\midrule
\endhead

\midrule
\multicolumn{5}{r}{\emph{(table continues on next page)}}\\
\bottomrule
\endfoot

\bottomrule
\endlastfoot

Const.           & \num{-0.04}          & \num{118.74}\sym{***} & \num{40.90}\sym{***} & \num{-0.18}\sym{***} \\
                 & \multicolumn{1}{c}{\small(0.10)}
                 & \multicolumn{1}{c}{\small(5.11)}
                 & \multicolumn{1}{c}{\small(2.50)}
                 & \multicolumn{1}{c}{\small(0.05)} \\

TIA              & \num{0.05}           & \num{1.98}            & \num{1.23}           & \num{0.05} \\
                 & \multicolumn{1}{c}{\small(0.10)}
                 & \multicolumn{1}{c}{\small(5.09)}
                 & \multicolumn{1}{c}{\small(2.55)}
                 & \multicolumn{1}{c}{\small(0.07)} \\

Apprec.          & \num{-0.05}          & \num{-2.38}           & \num{-1.07}          & \num{0.02} \\
                 & \multicolumn{1}{c}{\small(0.10)}
                 & \multicolumn{1}{c}{\small(5.09)}
                 & \multicolumn{1}{c}{\small(2.55)}
                 & \multicolumn{1}{c}{\small(0.07)} \\

Female           & \num{0.13}           & \num{1.56}            & \num{5.15}           & \num{0.05} \\
                 & \multicolumn{1}{c}{\small(0.15)}
                 & \multicolumn{1}{c}{\small(7.37)}
                 & \multicolumn{1}{c}{\small(3.60)}
                 & \multicolumn{1}{c}{\small(0.07)} \\

TIA $\times$ Female      & \num{0.00}   & \num{0.78}            & \num{-0.35}          & \num{-0.09} \\
                 & \multicolumn{1}{c}{\small(0.15)}
                 & \multicolumn{1}{c}{\small(7.34)}
                 & \multicolumn{1}{c}{\small(3.68)}
                 & \multicolumn{1}{c}{\small(0.10)} \\

Apprec. $\times$ Female  & \num{0.04}   & \num{1.34}            & \num{0.97}           & \num{-0.01} \\
                 & \multicolumn{1}{c}{\small(0.15)}
                 & \multicolumn{1}{c}{\small(7.34)}
                 & \multicolumn{1}{c}{\small(3.68)}
                 & \multicolumn{1}{c}{\small(0.10)} \\

ICC              & \num{0.50}           & \num{0.50}            & \num{0.50}           & \num{0.10} \\
\addlinespace[2pt]

\multicolumn{5}{@{}p{\dimexpr\linewidth-2\tabcolsep}@{}}{\footnotesize
\textit{Note:} Regression results with coefficients; standard errors in parentheses.
Baseline = Male. ${}^{*}p<0.10$, ${}^{**}p<0.05$, ${}^{***}p<0.01$.}\\

\end{longtable}